\documentclass[12pt]{article}

\pdfoutput=1

\usepackage{graphics}
\usepackage{epsfig}

\usepackage{amsfonts}
\usepackage{amssymb}
\usepackage{amsmath}
\usepackage{dsfont}
\usepackage{pifont}
\usepackage{bbm}
\usepackage{multirow}
\usepackage{latexsym}
\usepackage{verbatim}
\usepackage{mcite}
\usepackage{cite}
\usepackage{units}
\usepackage[all]{xy}
\usepackage{cancel}
\usepackage{slashed}
\usepackage{textcomp}

\def\beq{\begin{equation}}
\def\eeq{\end{equation}}
\def\beqa{\begin{eqnarray}}
\def\eeqa{\end{eqnarray}}

\def\g{{\gamma}}

\def\m{{\mu}}

\def\bfone{\relax{\rm 1\kern-.35em 1}}

\newcommand{\be}{\begin{equation}}
\newcommand{\ee}{\end{equation}}
\newcommand{\ben}{\begin{displaymath}}
\newcommand{\een}{\end{displaymath}}
\newcommand{\bea}{\begin{eqnarray}}
\newcommand{\eea}{\end{eqnarray}}

\newcommand{\bean}{\begin{eqnarray*}}
\newcommand{\eean}{\end{eqnarray*}}

\DeclareMathAlphabet{\mathpzc}{OT1}{pzc}{m}{it}

\begin{document}
\pagestyle{plain}


\makeatletter \@addtoreset{equation}{section} \makeatother
\renewcommand{\thesection}{\arabic{section}}
\renewcommand{\thefootnote}{\arabic{footnote}}
\renewcommand{\theequation}{\thesection.\arabic{equation}}


\setcounter{page}{1} \setcounter{footnote}{0}


\begin{titlepage}
\begin{flushright}
\small ~~
\end{flushright}

\bigskip

\begin{center}

\vskip 0cm

{\LARGE \bf  6d surface defects from massive type IIA} \\[6mm]

\vskip 0.5cm

{\bf Giuseppe Dibitetto$^1$ \,and\, Nicol\`o Petri$^2$}\let\thefootnote\relax\footnote{giuseppe.dibitetto@physics.uu.se, nicolo.petri@mi.infn.it}\\

\vskip 0.5cm

{\em 
$^1$Institutionen f\"or fysik och astronomi, University of Uppsala, \\ Box 803, SE-751 08 Uppsala, Sweden \\
$^2$Dipartimento di Fisica, Universit\`a di Milano, and INFN, Sezione di Milano,\\ Via Celoria 16, I-20133 Milano, Italy}

\vskip 0.8cm

\end{center}

\vskip 1cm

\begin{center}

{\bf ABSTRACT}\\[3ex]

\begin{minipage}{13cm}
\small

We present a new BPS flow within minimal $\mathcal{N}=1$ supergravity in seven dimensions describing a warped $\textrm{AdS}_{3}$ background supported by a ``dyonic'' profile of the three-form.
Furthermore, we discuss the holographic interpretation of the above solution in terms of a defect $\textrm{SCFT}_{2}$ inside the 6d $(1,0)$ theory dual to the AdS in the asymptotic region.
Finally we provide the brane picture of the aforementioned defect CFT as D2- and wrapped D4-branes ending on a D6 -- NS5 -- D8 funnel in massive type IIA string theory. 

\end{minipage}

\end{center}

\vfill

\end{titlepage}


\tableofcontents


\section{Introduction}

Conformal field theories (CFT) emerge in various different contexts in physics as effective descriptions of quantum critical systems, ranging from superconductors to black holes.
Generically, within the QFT framework, CFT's may appear as fixed points of renormalization group (RG) flows and hence potentially carry extremely valuable information concerning physics at strong 
coupling. In particular, in some cases the physics of the fixed point is even characterized by the loss of a Lagrangian description. 

Ever since the first brane constructions in string theory giving rise to SCFT's were given, our understanding of field theories has undergone a radical change. In particular, the discovery of interacting 
SCFT's in dimension higher than four in the context of string theory imposed a paradigm shift in the very way we think about QFT. Focusing on six-dimensional theories, their main peculiarity is that of
generically lacking a Lagrangian description, despite enjoying the property of locality. It is worth mentioning that, while the majority of stringy examples of interacting 6d CFT's are supersymmetric, 
the existence of non-supersymmetric examples has so far only been hinted at\footnote{We refer \emph{e.g.} to the construction in \cite{Apruzzi:2016rny}, employing $\overline{\textrm{D}6}$-branes in massive
type IIA backgrounds.}, and is still awaiting confirmations. 

Within the plethora of SCFT constructions obtained by taking the decoupling limit of string, M- or F-theory, the most known example of 6d SCFT is the $(2,0)$ theory living on the worldvolume of a stack of
coincident M5-branes \cite{Strominger:1995ac}. More in general, the range of all possible maximally supersymmetric conformal theories is exhausted by the so-called ADE classification, which can be 
accessed through geometric engineering in type IIB string theory \cite{Witten:1995zh}. When decreasing the amount of supersymmetry down to $(1,0)$, there appears to be a much richer structure.
The recent interesting proposals of \cite{Heckman:2013pva,DelZotto:2014hpa,Heckman:2015bfa} represent substantial progress towards a complete classification.

The constructions yielding $(1,0)$ theories which are relevant for our purposes were carried out in the context of massive type IIA by employing NS5 -- D6 -- D8 brane systems 
\cite{Hanany:1997gh,Brunner:1997gf,Brunner:1997gk}. According to the analysis performed by the work of \cite{Apruzzi:2013yva,Gaiotto:2014lca,Apruzzi:2014qva,Apruzzi:2015zna} over the last few years, 
all of the field theory models obtained from the above brane intersection admit a holographic description in terms of supersymmetric $\textrm{AdS}_{7}\times S^{3}$ solutions of massive type IIA 
supergravity. In the above references, firstly an exhaustive classification of all supersymmetric $\textrm{AdS}_{7}$ solutions was developed and secondly their holographic interpretation was proposed. 

When moving to the understanding of the dynamical process through which fixed points are reached, the issue that becomes of utmost importance is that of relating RG flows to brane movements in the underlying
brane picture of a given field theory. At the level of the gravity dual, such a process is captured by supergravity interpolating solutions known as domain walls (DW). This fact is sometimes referred to as
the DW/QFT correspondence \cite{Boonstra:1998mp}. In this holographic description of RG flows, the scalar fields which assume a non-trivial profile are identified with relevant operators realizing a 
marginal deformation of the original CFT, thus triggering the flow. The space of all exactly marginal deformations is called the conformal manifold.
However, in the case of minimal supersymmetry in 6d, the conformal manifold has been recently proven to be \emph{empty} \cite{Cordova:2015fha,Louis:2015mka}.

This opens up the alternative possibility for 6d SCFT's to admit a \emph{flow across dimensions} instead. In this case, the theory flows to a lower-dimensional fixed point which emerges as an effective description in the IR.
On the gravity side, this can be depicted as a different type of interpolating solution separating two AdS vacua in different dimensions. A stringy interpretation of this within the brane picture is 
usually given by spontaneous wrapping of branes in the presence of non-contractible cycles inside of their worldvolume. 
Some very well-known examples include spontaneous wrapping of M5-branes describing flows from the $(2,0)$ UV theory to 4d, 3d or 2d IR conformal fixed points \cite{Maldacena:2000mw,Gauntlett:2006ux,Gaiotto:2009gz,Bah:2012dg,Benini:2013cda}.
Some more recent and exotic examples of such flows involving brane wrapping in massive type IIA on punctured Riemann surfaces may be found in \cite{Bah:2017wxp}.

A somewhat complementary picture to the above one, is that of viewing a lower-dimensional CFT as the theory living on a \emph{conformal defect} inducing a position-dependent coupling in the ``mother CFT''.
The typical signature of the breaking of higher-dimensional conformal symmetry induced by the defect is the presence of non-vanishing one-point correlators as well as a displacement operator associated
with a non-conserved energy-momentum tensor. Starting from the seminal work of \cite{Karch:2000gx}, many stringy realizations of defect CFT's have been given in the literature. All of them rely on 
the study of boundary conditions of branes ending on other branes (see \emph{e.g.} the case of wrapped D5-branes ending on D3-branes, describing codimension 1 defects inside $\mathcal{N}=4$ SYM \cite{Gutperle:2012hy},
or M2- ending on M5-branes describing surface defects in the $(2,0)$ \cite{Lunin:2007ab}, or the more recent constructions by \cite{Maruyoshi:2016caf} of surface defects in class-$\mathcal{S}$ theories).
From the viewpoint of the supergravity dual description, such physical situations are described by asymptotically AdS flows involving a special slicing where the otherwise-flat slices are replaced by a
 lower-dimensional AdS space.

The goal of our paper is that of proposing a novel construction of conformal surface defects in the $(1,0)$ theories constructed from NS5 -- D6 -- D8 systems in massive type IIA string theory.
Our description is given in terms of D2- \& wrapped D4-branes ending on the above brane intersection. 
To this end, after a brief review of \cite{Hanany:1997gh}, we move to presenting a novel BPS flow of the above type in minimal 
$\mathcal{N}=1$ 7d supergravity with $\textrm{SU}(2)$ gauge group. It will feature an $\textrm{AdS}_{3}$ slicing supported by a non-trivial profile for the 3-form potential \cite{Dibitetto:2017tve}. 
Subsequently, we give its 10d lift and explicitly construct the brane intersection in massive type IIA supergravity containing both $\textrm{AdS}_{7}$ and $\textrm{AdS}_{3}$ in different limits.
Finally we conclude by sketching a computation of the one-point functions of the dual defect $\mathcal{N}=(0,4)$ 2d theory.

\section{6d $(1,0)$ SCFT's from massive type IIA}

Six-dimensional $\mathcal{N}=(1,0)$ QFT's enjoy eight real chiral supercharges transforming as a doublet of the R-symmetry group $\textrm{SU}(2)_{R}$. The standard vector multiplets (VM) can be coupled to extra
matter such as hypermultiplets (HM) and the more exotic tensor multiplets (TM). The bosonic field content of these latter ones comprizes a real scalar $\phi$ and a \emph{self-dual} 2-form field $b_{(2)}$.
The special branch of moduli space called the ``tensor branch'' is precisely parametrized by the vev of $\phi$.
The bosonic Lagrangian describing the coupling between the VM's and one Abelian TM is sketchily given by 
\be
\label{Lagrangian_6d}
\mathcal{L}_{\textrm{6d}} \, = \, \phi\,\textrm{Tr}\left(\left|F_{(2)}\right|^{2}\right)\,+\,\left(\partial\phi\right)^{2}\,+\,\left|db_{(2)}\right|^{2}\,+\,
*_{(6)}\left(b_{(2)}\,\wedge\,F_{(2)}\,\wedge\,F_{(2)}\right)\ ,
\ee
where the vev $\langle\phi\rangle$ parametrizing the tensor branch may be seen as the effective gauge coupling $g_{\textrm{YM}}^{-2}$.
From this perspective, it appears clear that the singular point $\langle\phi\rangle\,=\,0$ is related to physics at strong coupling, even though taking such a limit cannot be done naively since it involves
some subtleties. The valuable lesson that we learned from stringy constructions of supersymmetric QFT's is that this strongly coupled regime corresponds to a fixed point of an RG flow described by an 
interacting SCFT \cite{Seiberg:1996vs}.

Though several constructions in string and M-theory yielding $(1,0)$ SCFT's are available in the literature, there is still no exhaustive classification. However, a significant step in this direction has
been recently taken in \cite{Chang:2017xmr} by applying conformal bootstrap techniques. 

Since the aim of this paper is that of providing a novel holographic description of surface defects within $\mathcal{N}=(1,0)$ SCFT's, let us move further to the details of the stringy construction in which
such theories naturally arise. This will be the setting upon which our proposal relies. 
The original construction proposed in \cite{Hanany:1997gh} in the context of massive type IIA string theory \cite{Romans:1985tz} realizes a class of linear quivers which may be regarded as the six-dimensional analog of the 
Hanany-Witten constructions of \cite{Hanany:1996ie} obtained in a three-dimensional case. The brane system underlying these 6d field theories is made of NS5-, D6- and D8-branes, placed such in a way as
to preserve eight real supercharges, as shown in table~\ref{Table:GT}.
\begin{table}[h!]
\renewcommand{\arraystretch}{1}
\begin{center}
\scalebox{1}[1]{
\begin{tabular}{c||c c c c c c|c||c c c}
branes & $t$ & $y^{1}$ & $y^{2}$ & $y^{3}$ & $y^{4}$ & $y^{5}$ & $z$ & $r$ & $\theta^{1}$ & $\theta^{2}$ \\
\hline \hline
NS5 & $\times$ & $\times$ & $\times$ & $\times$ & $\times$ & $\times$ & $-$ & $-$ & $-$ & $-$ \\
D6 & $\times$ & $\times$ & $\times$ & $\times$ & $\times$ & $\times$ & $\times$ & $-$ & $-$ & $-$ \\
D8 & $\times$ & $\times$ & $\times$ & $\times$ & $\times$ & $\times$ & $-$ & $\times$ & $\times$ & $\times$ \\
\end{tabular}
}
\end{center}
\caption{{\it The brane picture underlying the 6d $(1,0)$ SCFT described by a NS5 -- D6 -- D8 system. The above system is $\frac{1}{4}$ -- BPS.
Note that the radial coordinate realizing the dual $\textrm{AdS}_{7}$ geometry turns out to be a combination of $z$ \& $r$.}} \label{Table:GT}
\end{table}

The most general linear quiver arising as the low-energy description of the brane system in table~\ref{Table:GT} contains $N$ VM's which stem from D6-brane worldvolume dynamics (and giving rise to the 
$\textrm{SU}(N)$ gauge symmetry), HM's in the fundamental representation due to the presence of D8-branes, and finally a bi-fundamental HM's for each NS5 and a TM for each pair of NS5's. 
In the stringy picture, the real scalar in the TM represents the relative distance between the two NS5-branes, which happens to be \emph{finite} in the tensor branch. Subsequently, the fixed point is
reached in the limit where the NS5-branes collide and tensionless string states appear in the spectrum of the effective field theory, which is therefore conformal.
This situation is depicted in figure~\ref{fig:(1,0)quiver}.
\begin{figure}[h]
\begin{center}
\scalebox{1}[1]{
\includegraphics[width=120mm]{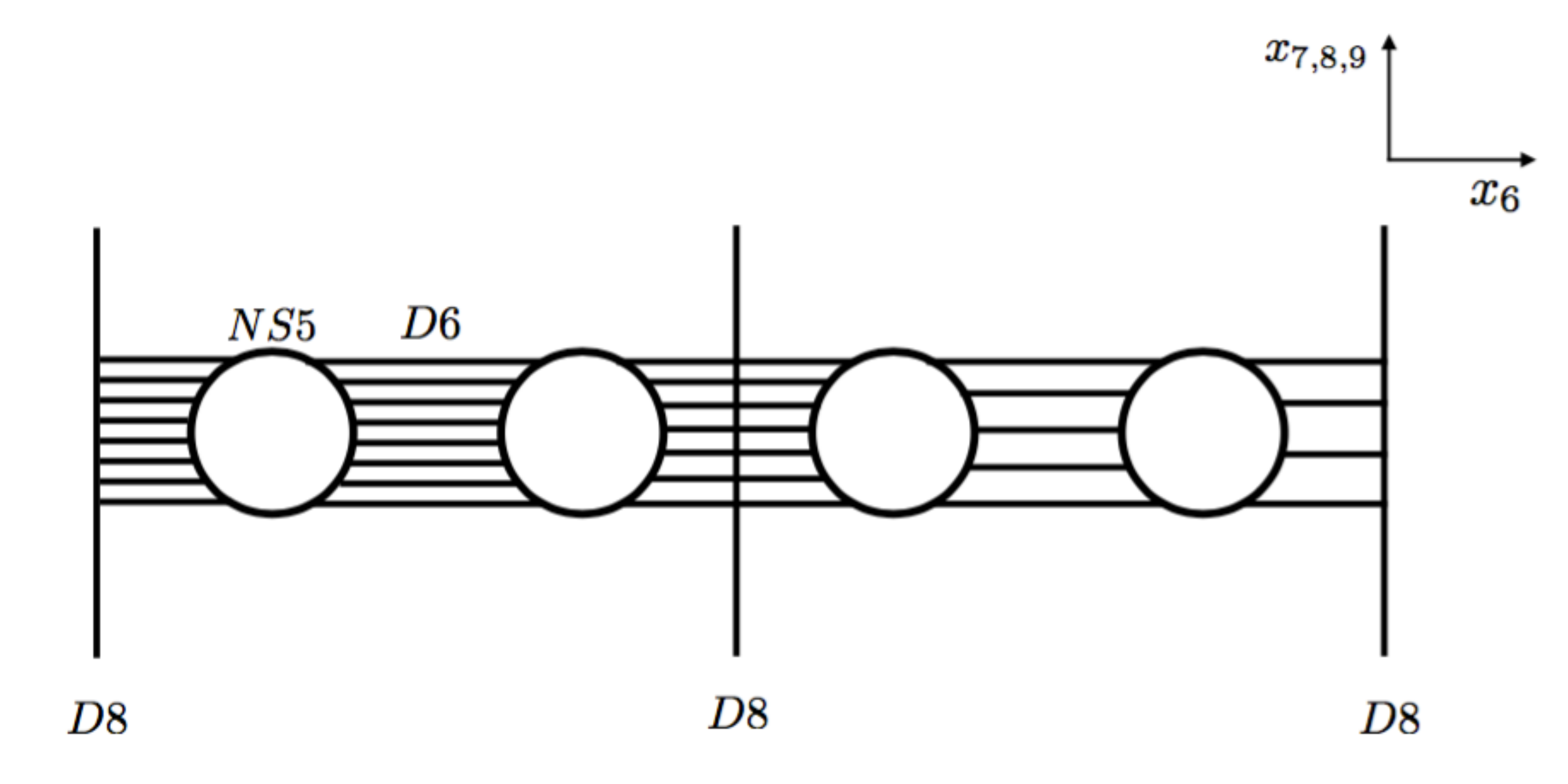}} 
\caption{\it The brane picture of the linear quiver realizing the above $(1,0)$ SCFT's. In this diagram the $x^{6}$ coordinate represents $z$, while the directions $x^{7,8,9}$ are to be identified with 
$(r,\theta^{1},\theta^{2})$. D8-branes are vertical lines, D6-branes are horizontal lines, and finally the ``fat'' bubbles represent point-like NS5-branes which collide at the fixed point.}
\label{fig:(1,0)quiver}
\end{center}
\end{figure}

As originally conjectured and motivated in \cite{Gaiotto:2014lca}, by making use of the AdS/CFT correspondence, the above linear quivers admit $\textrm{AdS}_{7}$ duals which were originally found in 
\cite{Apruzzi:2013yva} as BPS solutions of massive type IIA supergravity by using the pure spinor formalism (see also \cite{Apruzzi:2014qva,Apruzzi:2015zna,Passias:2016fkm} for further details).
These solutions stem from a warped compactification on a squashed $S^{3}$ obtained from a fibration of a round $S^{2}$ over a segment. In this gravity dual description D6-branes fill $\textrm{AdS}_{7}$
and are completely localized at the poles of $S^{3}$, while D8-branes wrap $S^{2}$ equators with finite volume. The $\textrm{SU}(2)_{R}$ R-symmetry group emerges here as the unbroken isometry group of 
the $S^{2}$.
In the original work of \cite{Gaiotto:2014lca} the relation of these solutions to a near-horizon limit of the above brane system was hinted at and later further clarified in \cite{Bobev:2016phc}.

\section{A new BPS flow in minimal $\mathcal{N}=1$ 7d supergravity}

Motivated by holographic reasons, we want to consider minimal $\mathcal{N}=1$ 7d supergravity with $\textrm{SU}(2)$ gauge group and non-zero topological mass. This theory is known to possess a 
supersymmetric $\textrm{AdS}_{7}$ vacuum preserving sixteen real supercharges \cite{Dibitetto:2015bia} and precisely admitting the massive type IIA interpretation given in the previous section. 
The field content of the theory in its minimal incarnation features the vielbein, an $\textrm{SU}(2)_{R}$-triplet of vector fields, a three-form gauge potential $B_{(3)}$ and a scalar field $X$, while
the fermionic degrees of freedom consist of symplectic-Majorana gravitini and dilatini, arranged into $\textrm{SU}(2)_{R}$-doublets.

In this paper we will restrict ourselves to the case of vanishing profile for the vectors. In this case, the bosonic Lagrangian reads \cite{Townsend:1983kk}
\be
\label{Lagrangian_7d}
\mathcal{L}_{\textrm{7d}} \, = \, \mathcal{R}\,*_{(7)}1\,-\,5\,X^{-2}\,*_{(7)}dX\,\wedge\,dX \,-\,\dfrac{1}{2}\,X^{4}\,*_{(7)}\mathcal F_{(4)}\,\wedge\, \mathcal F_{(4)} \,-\, \mathcal{V}(X)\,*_{(7)}1 \,-\, h\,\mathcal F_{(4)}\,\wedge\,B_{(3)}\ ,
\ee
where $\mathcal{R}$ denotes the 7d Ricci scalar, $\mathcal{V}(X)$ is the scalar potential and $\mathcal{F}_{(4)}$ is the field strength of the 3-form gauge potential.  
The explicit form of the scalar potential induced by a combination of gaugings and massive deformations reads
\be
\mathcal{V}(X) \ = \ 2h^{2} \, X^{-8} \, - \, 4\sqrt{2}\,gh\, X^{-3} \, - \, 2g^{2} \,X^{2} \ ,
\ee
where $g$ denotes the gauge coupling and $h$ the topological mass for the three-form field.
The above scalar potential can be rewritten in terms of a real \emph{superpotential} 
\be
\label{superpotential}
f(X) \ = \ \frac{1}{2}\,\left(h\,X^{-4}\,+\,\sqrt{2}\,g\,X\right) \ ,
\ee
through the relation
\be
\mathcal{V}(X) \ = \ \frac{4}{5}\,\left(-6f(X)^{2}\,+\,X^{2}\,\left(D_{X}f\right)^{2}\right) \ .
\ee
Finally, due to the presence of the topological term in \eqref{Lagrangian_7d} induced by $h$ and $B_{(3)}$, one has to impose an odd-dimensional self-duality condition \cite{Townsend:1983xs} of the form
\be
\label{SD_cond}
X^{4} \, *_{(7)}\mathcal{F}_{(4)}\ \overset{!}{=} \ -2h \, B_{(3)}  \ .
\ee

This supergravity theory enjoys $\mathcal{N}=1$ supersymmetry, which can be made manifest by checking the invariance of its Lagrangian w.r.t. the following supersymmetry transformations
\be
\label{SUSY_eqns_7D}
\left\{
\begin{array}{lclc}
\delta_{\epsilon}{\psi_{\m}}^{a} & = & \nabla_{\m}\epsilon^{a} \, + \,   
\,\frac{X^{2}}{160}\,\left({\g_{\m}}^{mnpq}\,-\,\frac{8}{3}\,{e_{\m}}^{m}\,\g^{npq}\right) \, \mathcal{F}_{(4)\, mnpq}\,\epsilon^{a} \, - \, \frac{1}{5}\,f(X)\,\g_{\m}\,\epsilon^{a} & , \\[2mm]
\delta_{\epsilon}{\chi^{a}} & = &  \frac{\sqrt{5}}{2}\,X^{-1}\slashed{\partial}X \, \epsilon^{a}  
\, + \, \frac{X^{2}}{2\sqrt{5}}\,  \slashed{\mathcal{F}}_{(4)}\,\epsilon^{a} \, - \, \frac{X}{5} \, D_{X}f \, \epsilon^{a} & ,
\end{array}\right.
\ee
where we introduced the following notation $\slashed{\omega}_{(p)}\,\equiv\,\frac{1}{p!}\,\g^{m_{1}\cdots m_{p}}\,\omega_{(p)\,m_{1}\cdots m_{p}}$, $\omega_{(p)}$ being a $p$-form.

\subsection*{The $\frac{1}{2}$ -- BPS solution with $3$-form profile}

Our present goal will be that of giving a holographic description of conformal defects within the 6d $(1,0)$ theories which are dual to the aforementioned AdS vacua. 
Following the logic in \cite{Clark:2004sb}, this in turn implies that we need to look for Janus-type solutions to the effective supergravity model in question. These solutions describe a foliation of
spacetime in terms of lower-dimensional AdS space slices. In our setup, the novel resulting solution will be asymptotically locally $\textrm{AdS}_{7}$ and will be furthermore characterized by an
$\textrm{AdS}_{3}\times S^{3}$ foliation.

Let us then cast the following \emph{Ansatz} for the 7d fields
\be
\begin{array}{lcll}
ds_{7}^{2} & = & e^{2U(r)}\,\left(ds_{\textrm{AdS}_{3}}^{2} \, + \,d\Omega_{(3)}^{2} \right)\, + \, e^{2V(r)} \, dr^{2} & , \\[2mm]
X & = & X(r) & , \\[2mm]
B_{(3)} & = & k(r)\,\left(\textrm{vol}_{(1,2)} \, + \, \textrm{vol}_{(3)}\right) & ,
\end{array}
\ee
where $ds_{\textrm{AdS}_{3}}^{2}$ \& $d\Omega_{(3)}^{2}$ respectively denote the unit $\textrm{AdS}_{3}$ \& the unit $S^{3}$ metric,
while $\textrm{vol}_{(1,2)}$ \& $\textrm{vol}_{(3)}$ represent the corresponding volume forms. 
For a suitable Killing spinor with a radial profile, the susy equations \eqref{SUSY_eqns_7D} are fully implied by the following first-order flow equations
\be\left\{
\begin{array}{lclclc}
U' & = & \frac{2}{5} \, e^{V} \, f  & , \\[2mm]
k' & = & -\frac{e^{2 U+V}}{X^2} & , \\[2mm] 
X' & = & -\frac{2}{5} \, e^{V} \, X^{2} \,  D_{X}f & ,
\end{array}\right.
\ee
where $f$ denotes the superpotential introduced in \eqref{superpotential}, where we furthermore fix $h\equiv\,\frac{g}{2\sqrt{2}}$ so as to have the susy AdS vacuum at $X=1$.

After performing the following gauge choice for the function $V$
\be
e^{-V} \, \overset{\textrm{gauge fix.}}{=} \, -\frac{2}{5} \, e^{V} \, X^{2} \,  D_{X}f \ ,
\ee
the above flow equations may be integrated analytically and the solution reads
\be
\label{7dSOL}
\begin{array}{lclclclclc}
e^{2U} & = & \dfrac{2^{-1/4}}{\sqrt{g}}\,\sqrt{\dfrac{r}{1\,-\,r^{5}}} & & , & & e^{2V} & = & \dfrac{25}{2g^{2}}\,\dfrac{r^{6}}{\left(1\,-\,r^{5}\right)^{2}} & , \\[3mm]
k & = & -\dfrac{2^{1/4}}{g^{3/2}}\sqrt{\dfrac{r^{5}}{1\,-\,r^{5}}} & & , & & X & = & r & ,
\end{array}
\ee
where $r$ ranges from $0$ to $1$.
One may check that \eqref{7dSOL} correctly satisfies the bosonic field equations as well as the odd-dimensional self-duality condition \eqref{SD_cond}. Note that this solution is asymptotically 
locally $\textrm{AdS}_{7}$ since, as $r\,\rightarrow\,1^{-}$,
\be
\begin{array}{lclc}
X & = & 1 \, + \, \mathcal{O}(1-r) & , \\[2mm]
\mathcal{R}_{7} & = & -\frac{21}{4}\,g^{2} \, + \, \mathcal{O}\left((1-r)^{2}\right) & ,
\end{array}
\ee
while as $r\,\rightarrow\,0$ it has an IR singularity of the form
\be
\label{IR_sing}
\begin{array}{lclclclclc}
e^{2U} & = & \dfrac{r^{1/2}}{2^{1/4}\sqrt{g}}\, + \, \mathcal{O}(r^{5/2}) & & , & & e^{2V} & = & \dfrac{25r^{6}}{2g^{2}}\, + \, \mathcal{O}(r^{7}) & , \\[3mm]
k & = & -\dfrac{2^{1/4}}{g^{3/2}}\,r^{5/2} \, + \, \mathcal{O}(r^{7/2}) & & , & & X & = & r & .
\end{array}
\ee
In the next section we will see how the above IR singularity can be interpreted as a brane singularity in 10d. In particular, we will see how the \emph{dyonic} $\mathcal{F}_{(4)}$ singularity is related to
D2- and D4-branes filling $\textrm{AdS}_{3}$.
Generically, away from the two above limits, this solution describes a novel 7d supersymmetric background obtained as warped product of $\textrm{AdS}_{3}$ times a 4d hyperbolic space constructed as a 
fibration of $S^{3}$ over a segment.

\subsection*{Massive type IIA lift}

Minimal $\textrm{SU}(2)$ gauged 7d $\mathcal{N}=1$ supergravity is known to arise from a consistent truncation of massive type IIA supergravity on a squashed $S^{3}$ \cite{Passias:2015gya}.
By employing the uplift formulae there, we find
\be
\label{brane_sol}
\begin{array}{lclc}
ds_{10}^{2} & = & \frac{\sqrt{2}}{g} \, X^{-1/2} \, e^{2A} \, \left(e^{2U}\,\left(ds_{\textrm{AdS}_{3}}^{2} +d\Omega_{(3)}^{2} \right)\, + \, e^{2V} \, dr^{2}\right) 
\, + \, X^{5/2} \, \left(dy^{2}\,+\,\frac{1-x^{2}}{16 w}e^{2A}\,d\Omega_{(2)}^{2}\right)& , \\[3mm]
e^{2\Phi} & = & \frac{8\sqrt{2}}{g^{3}}\,\frac{X^{5/2}}{w}\,e^{2\Phi_{0}} & , \\[3mm]
B_{(2)} & = & \frac{1}{\sqrt{2}g^{3}}\,e^{2A}\,\frac{x\sqrt{1-x^{2}}}{w}\,\textrm{vol}_{(2)} 
\,-\, \frac{4\sqrt{2}}{g^{3}}\,e^A\,dy\,\wedge\,\psi & , \\[3mm]
F_{(0)} & = & m & , \\[3mm]
F_{(2)} & = & e^{A-\Phi_{0}}\,\sqrt{1-x^{2}}\,\left(-\frac{1}{4}\,+\,\frac{m}{\sqrt{2}g^{3}\,w}\,e^{A+\Phi_{0}}\,x\right)\,\textrm{vol}_{(2)} & , \\[3mm]
F_{(4)} & = & -e^{2A-\Phi_{0}}\,\left(\frac{4\sqrt{2}}{g}X^{4}\,\sqrt{1-x^{2}}\,dy\,\wedge\,*_{(7)}\mathcal{F}_{(4)}\,+\,\frac{x}{2}\,e^{A}\,\mathcal{F}_{(4)}\right) & , 
\end{array}
\ee
with $w\,\equiv\,X^{5}\,\left(1-x^{2}\right)\,+\,x^{2}$, and the 1-form $\psi$ is such that $-2\,d\psi\,=\,\textrm{vol}_{(2)}$.
In the above 10d solution $U$, $V$, $X$ \& $\mathcal{F}_{(4)}$ are the radial functions given in \eqref{7dSOL}, while $A$, $x$ \& $\Phi_{0}$ are functions of the $y$ coordinate satisfying the following
first-order flow equations
\be\label{eq:oder}
\left\{
\begin{array}{lclc} 
\frac{d\Phi_{0}}{dy} & = & \dfrac{1}{4} \dfrac{e^{-A}}{\sqrt{1-x^2}} \, \left(12 \, x \, + \, \left(2x^2-5\right)\,m \, e^{A+\Phi_{0}}\right) & , \\[4mm]
\frac{dx}{dy} & = & -\dfrac{1}{2} e^{-A}\sqrt{1-x^2} \, \left(4 \, + \, x \, m\, e^{A+\Phi_{0}}\right) & ,\\[3mm]
\frac{dA}{dy} & = & \dfrac{1}{4} \dfrac{e^{-A}}{\sqrt{1-x^2}} \, \left(4\,x \, - \, m \, e^{A+\Phi_{0}}\right) & ,
\end{array}\right.
\ee
and thus completely specifying the warping. 

\section{Massive type IIA brane picture}

The brane construction underlying the physical system studied in this paper can be understood in two steps. The first part of the construction is the one setting up the ``boundary CFT'', \emph{i.e.} the 
$(1,0)$ theory and it is made of the D6 -- NS5 -- D8 funnel as explained in the first section. 
The corresponding massive type IIA supergravity background can be described in terms of a special class of $\frac{1}{4}$ -- BPS flows as those introduced in \cite{Imamura:2001cr}.

The second part of our construction is here given by D2-branes and wrapped D4-branes ending on the above massive brane funnel. The physics of the boundary conditions of these branes in the context of
massive type IIA string theory turns out to be described by a $\textrm{SCFT}_{2}$ living on a codimension 4 defect inside the original 6d spacetime.

The complete brane system realizing this mechanism is sketched in table~\ref{Table:branes}.
\begin{table}[h!]
\renewcommand{\arraystretch}{1}
\begin{center}
\scalebox{1}[1]{
\begin{tabular}{c||c c|c c c c|c||c c c}
branes & $t$ & $y$ & $\rho$ & $\varphi^{1}$ & $\varphi^{2}$ & $\varphi^{3}$ & $z$ & $r$ & $\theta^{1}$ & $\theta^{2}$ \\
\hline \hline
NS5 & $\times$ & $\times$ & $\times$ & $\times$ & $\times$ & $\times$ & $-$ & $-$ & $-$ & $-$ \\
D6 & $\times$ & $\times$ & $\times$ & $\times$ & $\times$ & $\times$ & $\times$ & $-$ & $-$ & $-$ \\
D8 & $\times$ & $\times$ & $\times$ & $\times$ & $\times$ & $\times$ & $-$ & $\times$ & $\times$ & $\times$ \\
\hline
D2 & $\times$ & $\times$ & $-$ & $-$ & $-$ & $-$ & $\times$ & $-$ & $-$ & $-$ \\
D4 & $\times$ & $\times$ & $-$ & $-$ & $-$ & $-$ & $-$ & $\times$ & $\times$ & $\times$ \\
\end{tabular}
}
\end{center}
\caption{{\it The brane picture underlying the 2d SCFT described by D2- and D4-branes ending on an NS5 -- D6 -- D8 funnel. The above system is $\frac{1}{8}$ -- BPS.}} \label{Table:branes}
\end{table}

The 10d supergravity background corresponding to the brane system sketched in table~\ref{Table:branes} can be constructed as a non-harmonic superposition of a solution from \cite{Imamura:2001cr} and 
D2- \& D4-branes, yielding
\be
\label{brane_sol}
\begin{array}{lclc}
ds_{10}^{2} & = & S^{-1/2}H_{\textrm{D}2}^{-1/2}H_{\textrm{D}4}^{-1/2}\,ds_{\textrm{Mkw}_{2}}^{2}\,+\, 
S^{-1/2}H_{\textrm{D}2}^{1/2}H_{\textrm{D}4}^{1/2}\,\left(d\rho^{2}+\rho^{2}\,d\Omega_{(3)}^{2}\right) \,+ & \\[2mm]
& + & K\,S^{-1/2}H_{\textrm{D}2}^{-1/2}H_{\textrm{D}4}^{1/2}\,dz^{2}\,+\,K\,S^{1/2}H_{\textrm{D}2}^{1/2}H_{\textrm{D}4}^{-1/2}\,\left(dr^{2}+r^{2}\,d\Omega_{(2)}^{2}\right) & , \\[3mm]
e^{\Phi} & = & g_{s}\,K^{1/2}\,S^{-3/4}H_{\textrm{D}2}^{1/4}H_{\textrm{D}4}^{-1/4} & , \\[3mm]
H_{(3)} & = & \frac{\partial}{\partial z}\left(KS\right)\textrm{vol}_{(3)}\,-\,dz\,\wedge\,*_{(3)}\left(dK\right) & , \\[3mm]
F_{(0)} & = & m & , \\[3mm]
F_{(2)} & = & -g_{s}^{-1}\,*_{(3)}\left(dS\right) & , \\[3mm]
F_{(4)} & = & g_{s}^{-1}\,\textrm{vol}_{(1,1)}\,\wedge\,dz\,\wedge\,dH_{\textrm{D}2}^{-1} \, + \, 
*_{(10)}\left(\textrm{vol}_{(1,1)}\,\wedge\,\textrm{vol}_{(3)}\,\wedge\,dH_{\textrm{D}4}^{-1}\right) & , 
\end{array}
\ee
where the functions $K(z,r)$ \& $S(z,r)$ satsify \cite{Imamura:2001cr}
\be
\left\{
\begin{array}{rccc}
mg_{s}\,K \,-\, \frac{\partial S}{\partial z} & = & 0 & , \\[3mm]
\Delta_{(3)}S \, + \, \frac{1}{2}\frac{\partial^{2}}{\partial z^{2}} S^{2} & = & 0 & ,
\end{array}
\right.
\ee
while 
\be
\begin{array}{lccclc}
H_{\textrm{D}2}(\rho,r) \ = \ \left(1+\frac{q_{\textrm{D}4}}{\rho^{2}}\right)\left(1+\frac{q_{\textrm{D}6}}{r}\right) & , & & & 
H_{\textrm{D}4}(\rho) \ = \ \left(1+\frac{q_{\textrm{D}4}}{\rho^{2}}\right) & .
\end{array}
\ee

\begin{figure}
\begin{center}
\scalebox{1}[1]{\xymatrix{*+[F-,]{\textrm{AdS}_{7}\times_{\textrm{W}}\tilde{S}^{3}} & *+[F-,]{\begin{array}{c} \textrm{pure D8}\\ \textrm{limit}\end{array}} \\ 
*+[F-,]{\begin{array}{c} \textrm{D2 -- D4 -- D6 -- NS5 -- D8}\\ \textrm{bound state}\end{array}} \ar[r]_{\hspace{13mm}\rho}\ar[u]^{\zeta\,\equiv\,\left(r\,+\,z^{2}\right)^{-1/2}}\ar[ru]^{z} 
 &  *+[F-,]{\textrm{AdS}_{3}\times_{\textrm{W}}\mathcal{M}_{7}}
 }}
\end{center}
\caption{{\it The three different limits of the brane system represented in \eqref{brane_sol} depending on the three coordinates $(\rho,z,r)$, respectively yielding $\textrm{AdS}_{7}$, 
the asymptotic domain-wall behavior typical of massive type IIA solutions, and $\textrm{AdS}_{3}$. Each limit is controlled by a different combination of the above cooridnates.}}\label{fig:three_limits} 
\end{figure}
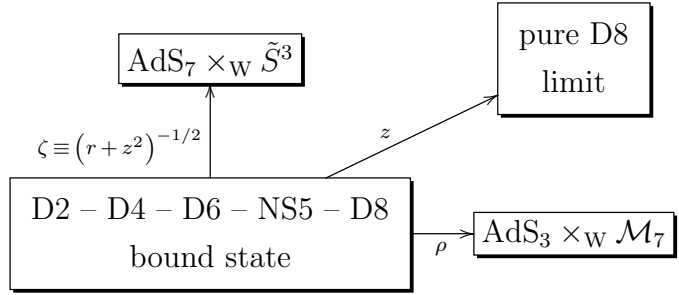

The above background may be regarded as a massive generalization of the special ``non-standard'' D2 -- D4 -- NS5 -- D6 intersection found in \cite{Boonstra:1998yu}.
\begin{itemize}
\item \textbf{AdS$_7$ from \eqref{brane_sol}:} Take $K\, \sim \, \frac{2}{z^{3}} \, G\left(\frac{r}{z^{2}}\right)$, and $S\, \sim \, \frac{1}{4r} \, H\left(\frac{r}{z^{2}}\right)$, for some 
suitable functions $G$ \& $H$. Now take $\rho\,\rightarrow\,\infty$ (which effectively gets rid of the D4-brane charge) and perform the following coordinate redefiniton (see \emph{e.g.} \cite{Cvetic:2000cj})
\be
\left\{
\begin{array}{lclc}
r^{1/2} & \equiv & \frac{\sin\alpha}{\zeta} & , \\[2mm]
z & \equiv & \frac{\cos\alpha}{\zeta} & ,
\end{array}
\right.
\ee
after which, upon choosing $H\,=\,1$ and $G\,=\,\frac{1}{2}\,\cos^3\alpha$, the metric in this limit reads
\be
ds_{10}^{2}\ \sim\ 2\,\cos\alpha \,\left(\tan\alpha\,ds_{\textrm{AdS}_{7}}^{2} \,+\,\tan\alpha\,d\alpha^{2}\,+\,
\frac{1}{4}\,\sin^{2}\alpha\,d\Omega_{(2)}^{2}\right) \ ,
\ee
which is nothing but $\textrm{AdS}_{7}\times_{\textrm{W}}\tilde{S}^{3}$, where $\tilde{S}^{3}$ is a 3-manifold topologically spherical and obtained as a fibration of a round $S^{2}$ over a segment.

\item \textbf{AdS$_3$ from \eqref{brane_sol}:} Now take $z$ \& $r$ to $\infty$ (while still keeping $\frac{r}{z^{2}}$ finite!) and send $\rho\,\rightarrow\,0$. In this limit, 10d metric will look like a warped 
product of an effective 7d metric and the above $\tilde{S}^{3}$, the warping being parametrized by the $\alpha$ coordinate.
If we focus on the 7d block of the metric, we find
\be
ds_{7}^{2}\ \sim \ \zeta^{-1/4}\,\underbrace{\left(\frac{\rho^{2}}{q_{\textrm{D}4}}\,ds_{\textrm{Mkw}_{2}}^{2}\,+\,\frac{q_{\textrm{D}4}}{\rho^{2}}\,d\rho^{2}\right)}_{ds_{\textrm{AdS}_{3}}^{2}}
\,+\,\frac{d\zeta^{2}}{\zeta^{2}}\,+\,q_{\textrm{D}4}\,\zeta^{-1/4}\,d\Omega_{(3)}^{2}\ ,
\ee
which is the warped product of $\textrm{AdS}_{3}$ with a 4-manifold $\mathcal{M}_{4}$ constructed as a fibration of $S^{3}$ over a segment.
\end{itemize}
To summarize, the brane bound state given in \eqref{brane_sol} comprizes three different limits that can be taken, two of which are holographically relevant. These respectively correspond to 
taking a near-horizon limit of the NS5 -- D6 -- D8 funnel to access the $(1,0)$ $\textrm{SCFT}_{6}$, and approaching the D2 -- D4 bound state to probe the defect $(0,4)$ $\textrm{SCFT}_{2}$. 
The third limit just describes how to move far away from all sources in the system to recover the correct domain-wall asymptotic behavior which is expected from the presence of D8-branes.
This situation is sketched in figure~\ref{fig:three_limits}. 

\section{One-point correlation functions}

To conclude, we would like to make use of the above holographic description in order to make a prediction concerning the one-point correlation functions of the $(1,0)$ theory in the presence of a surface
defect. Even in the case of a non-Lagrangian theory like ours, one can still think of the defect as some operator insertion realizing a deformation in the original SCFT.
Non-vanishing one-point correlators can be seen as a typical consequence of the fact that $\textrm{SO}(2,6)$ symmetry is broken by the aforementioned deformation.
Following the philosophy in \cite{Clark:2004sb}, we will sketch the computation by using two different methods: a holographic one which uses the standard holographic dictionary to extract the desired information 
from the gravity dual, and a field-theoretical one, which relies on the conformal symmetry preserved by the defect.
Of course, as opposed to \cite{Clark:2004sb}, due to a lack of a Lagrangian description, we will only be able to derive a matching constraint between unknown parameters on the two sides.

\subsection*{Holographic method}

Starting from the solution in equation~\eqref{7dSOL}, we first extract the boundary metric to get
\be
ds_{7}^{2}\ = \ F^{-2} \,\bigg(\underbrace{ds_{\textrm{Mkw}_{2}}^{2}\,+\,d\rho^{2}\,+\,\rho^{2}d\Omega_{(3)}^{2}}_{ds_{\textrm{Mkw}_{6}}^{2}}\,+\,\rho^{2}dR^{2}\bigg) \ ,
\ee
where $F\,\equiv\,e^{-U}\,\rho$ and the new coordinate $R$ has been introduced such that $dR\,\overset{!}{=}\,e^{V}\,dr$. One gets easily convinced that, by performing the different gauge choice 
$e^{V} \, \overset{\textrm{gauge fix.}}{=} \,1$, the metric in the $(\rho,R)$-plane has a conical defect where the ``angular'' coordinate $r$ ranges from $0$ to $1$, thus identifying an angular wedge.

Now pick the scalar $X$ as the responsible for the deformation driving the flow under consideration. Near the boundary, its normalized mass reads \cite{Apruzzi:2016rny}
\be
\begin{array}{lccclc}
m_{X}^{2}\,\ell^{2} & = & -8 & \overset{!}{=} & \Delta_{X}\,(\Delta_{X}-6) & ,
\end{array}\notag
\ee
whence $\Delta_{X}\,=\,4$. In terms of the $R$ coordinate its asymptotic bahavior is then given by
\be
X(R) \ \sim \ 1 \, - \, c\,e^{\frac{R}{\sqrt{2}g}} \ , \ \ \textrm{as } \ R\,\rightarrow\,-\infty \ ,
\ee
where $c$ is an arbitrary constant. Now using the holographic prescription we can relate the vev of $X$ to the one-point function of its dual operator $\mathcal{O}_{X}$ through
\be
X(R) \ = \ 1 \, - \, b\,\langle\mathcal{O}_{X}\rangle \, F^{\Delta_{X}} \, + \, \dots \ ,
\ee
whence $\langle\mathcal{O}_{X}\rangle\,=\,\frac{1}{5\sqrt{2}g\,b}\,\rho^{-4}$.

\subsection*{Conformal perturbation theory method}

An alternative way of computing correlators for a CFT deformed by some operator insertion is that of using conformal perturbation theory. 
Inspired by cases admitting a Lagrangian description, the aforementioned perturbation may be viewed as an extra term in the Lagrangian of the form $\gamma\,\phi\,\mathcal{O}_{X}$, where $\phi$ is our 
position-dependent coupling (mass dimension $2$), and $\gamma$ is a dimensionless quantity, which is usually related to the anomalous dimension of $\mathcal{O}_{X}$. The only difference here w.r.t. the aforementioned cases is that the overall 
constant appearing in front of the one-point correlation functions cannot be fixed by using this method since no coupling should be assumed small.
Based on typical field theory intuition, one can treat the above marginal deformation as a perturbation which appears an operator insertion inside corrections of $n$-point correlators as
\be
\hspace{-4mm}
\begin{array}{lcl}
\langle\mathcal{O}_{1}(x_{1})\cdots\mathcal{O}_{n}(x_{n})\rangle_{\textrm{def.}} & = & \langle\mathcal{O}_{1}(x_{1})\cdots\mathcal{O}_{n}(x_{n})\rangle_{0} \, + \,  \\[2mm]
 & & \gamma\,\int d^{6}z \,\phi(z)\,\langle\mathcal{O}_{1}(x_{1})\cdots\mathcal{O}_{n}(x_{n})\,\mathcal{O}_{X}(z)\rangle_{0} \, + \, \\[2mm]
 & & \frac{\gamma^{2}}{2!}\,\int d^{6}z\int d^{6}w \,\phi(z)\phi(w)\,\langle\mathcal{O}_{1}(x_{1})\cdots\mathcal{O}_{n}(x_{n})\,\mathcal{O}_{X}(z)\,\mathcal{O}_{X}(w)\rangle_{0} \, + \, \dots 
\end{array}\notag
\ee
By applying the above formula for a one-point function of $\mathcal{O}_{1}\,\equiv\,\mathcal{O}_{X}$, we find
\be
\langle\mathcal{O}_{X}(\rho)\rangle_{\textrm{def.}}\,=\,\underbrace{\langle\mathcal{O}_{X}(\rho)\rangle_{0}}_{0}
\,+\,\gamma\,\int d^{6}z\,\phi(z)\,\underbrace{\langle\mathcal{O}_{X}(\rho)\mathcal{O}_{X}(z)\rangle_{0}}_{\frac{a}{|\rho-z|^{8}}}\,+\,\dots
\ee
Now we pick coordinates such that $d^{6}z\,\rightarrow\,\rho'^{3}d\rho'd\Omega_{(3)}\,d^{2}z$, and we find that we need to have a position-dependent coupling behaving as $\phi(\rho)\,\sim\,\rho^{-2}$ in
order to match the gravity result. By making this assumption, we find
\be
\langle\mathcal{O}_{X}(\rho)\rangle_{\textrm{def.}}\,=\,\frac{\pi^{3}}{30}\,a\gamma\,\rho^{-4}\ ,
\ee
which exactly matches the holographic prediction, provided that $(5\sqrt{2}g\,b)^{-1}\,=\,\frac{\pi^{3}}{30}\,a\gamma$.

\section{Conclusions}

In this paper we studied a particular type of BPS flow in minimal $\mathcal{N}=1$ supergravity in seven dimensions involving a ``dyonic'' profile for the three-form gauge potential.
This novel solution is of a Janus type and describes an $\textrm{AdS}_{3}$ slicing of the 7d metric. When lifted to massive type IIA supergravity, it yields a warped $\textrm{AdS}_{3}$ background.
We interpret this geometry as the holographic description of conformal surface defects in a $(1,0)$ 6d SCFT.

To support this interpretation, we worked out the brane picture of the aforementioned solution and found that it is given by D2- \& D4-branes intersecting the Hanany-Zaffaroni NS5 -- D6 -- D8 system 
describing the $(1,0)$ 6d SCFT which is dual to the $\textrm{AdS}_{7}$ asymptotic geometry. Note that both D2- \& D4-branes are crucial in our construction in order to support a dyonic profile
of the 10d RR three-form, which directly reduces to the 7d one. 

In the last part of the paper we further speculate on possible holographic predictions concerning the defect SCFT$_{2}$ that can derived directly from the gravity side. 
Despite the lack of a Lagrangian description, we could find holographic evindence
for non-vanishing one-point correlators of a scalar operator of conformal dimension four. This may be regarded as the main signature of defects.

A possible avenue to be pursued in this context could be that of studying the effects on the energy-momentum tensor of the theory due to the presence of surface defects. In particular, it would be
very interesting to perform a holographic computation of the \emph{displacement operator} for our case, following the lines of \cite{Balakrishnan:2016ttg}.

To conclude, yet another very interesting thing to be further investigated may be found on the gravity side. In order to have a complete description of the surface defect, it would be of utmost importance
to find a more general flow depending on two coordinates and describing a $\textrm{Mkw}_{2}$ slicing of the 7d metric which would then realize the smooth interpolation between $\textrm{AdS}_{3}$ in the IR
limit and $\textrm{AdS}_{7}$ in the UV. Such a solution would not only provide the holographic description of the defect CFT, but also account for the dynamic RG flow through which the defect emerges. 

%
%

\section*{Acknowledgments}

We would like to thank Pietro Longhi, Joseph Minahan, Raul Pereira, Luigi Tizzano for very interesting and stimulating discussions. 
We further thank Luigi Tizzano for some valuable comments on a draft version of this manuscript. NP would like to thank
the members of the Department of Theoretical Physics at the Uppsala University for their kind hospitality while this work was being prepared.
The work of GD is supported by the Swedish Research Council (VR), and the G\"oran Gustafsson Foundation.

\newpage

%
%

\small

\bibliography{references}
\bibliographystyle{utphys}

\end{document}